\documentclass[a4paper]{revtex4}
\usepackage{graphicx}
\usepackage{natbib}
\usepackage{epsfig}

\begin{document}

\title{Skewed superstatistical distributions from a Langevin and Fokker-Planck
approach}

\author{Erik Van der Straeten and Christian Beck}

\affiliation{Queen Mary University of London, School of Mathematical Sciences,
Mile End Road, London E1 4NS, UK}

\begin{abstract}
The superstatistics concept is a useful statistical method
to describe
inhomogeneous complex systems for which a system parameter
$\beta$ fluctuates on a large spatio-temporal scale.
In this paper we analyze
a measured time series of wind speed
fluctuations and extract the superstatistical
distribution function $f(\beta)$ directly from the data.
We construct suitable Langevin and Fokker-Planck models
with a position dependent $\beta$-field and show
that they reduce to standard type of superstatistics
in the overdamped limit.
\end{abstract}

\maketitle

\section{Introduction}
The superstatistics concept introduced
in 2003 \cite{beck-cohen} provides a useful tool
to describe a large variety of complex systems
\citep{unicla,touchette,supergen,souza,chavanis,vignat,Plastino-x,jizba,
prl01,frank,hase,celia,abul-again,straeten-recent}.
The basic idea is to characterize the
complex system under consideration by a superposition of two
statistics,
one corresponding to ordinary statistical mechanics (on a mesoscopic level
modeled
e.g.\ by a
Langevin equation) and the other one corresponding to a slowly
varying parameter $\beta (\vec{x},t)$ of the system which can
be, but need not to be, an
inverse temperature.
There are many interesting applications
of superstatistical techniques. Recent work in this direction
includes train delay statistics \cite{briggs},
the distribution
of accelerations of test particles
in turbulent flows \cite{prl} and cancer survival statistics
\cite{chen}. Further applications are described
in
\citep{daniels,maya,reynolds,RMT,abul-magd3,porpo,rapisarda,kantz,cosmic,eco,
vdsbeck1,superscatter,ausloos,kings,laub,abe-thurner,queiros,abul-magd2,abc,
crooks,naudts,
straeten,hau}.

In this paper, as a working example, we study measured statistics
of wind speed fluctuations. For previous work in this direction,
see \cite{rapisarda, kantz}. We will extract the superstatistical
distribution function $f(\beta)$ out of our data set and
show that in good approximation it is an inverse gamma distribution.
As a theoretical model for these and other data
sets, we will study Langevin and Fokker-Planck
equations with a varying, position-dependent $\beta$. We will show
that in the overdamped limit these models reduce to standard type
of superstatistics. Skewed stationary distributions
arise naturally in these types of models
if the overdamped limit is not completely performed.

This paper is organized as follows. In section II we recall the basic
concepts of superstatistics, illustrated with the example of a linear
Langevin equation. In section III we apply our methods to measured
wind speed data, extracting various relevant distributions and parameters
out of the data set. In section IV we extend our theoretical approach by
studying
a Kramers-like equation with spatially varying $\beta$. We show that this model
reduces to standard type of superstatistics in the overdamped limit. An example
is studied in section V.
Then, in section VI we comment on the skewness
of the generated distributions. The last section gives a short discussion of the
results of this paper.

\section{Linear Langevin equation with time-dependent $\beta$}
Let us study as
a simple example the following
Langevin equation
\begin{eqnarray}
\frac d{dt}v(t)&=&-\gamma v(t)+\sqrt{\frac{2\gamma}m\frac1{\beta(t)}}g(t),
\end{eqnarray}
where $\gamma$ and $m$ are constants and $g(t)$ corresponds to Gaussian white
noise with unit variance.
$v(t)$ is a stochastic process that can be associated with the velocity
of a Brownian particle.
If $\beta(t)=\beta$ is a constant, then
one ends up with the standard Langevin equation which describes the dynamical
process of a Brownian particle with mass $m$ in an environment with constant
temperature $1/\beta$. It is well known that the stationary velocity
distribution of a Brownian particle is a Gaussian distribution with zero
mean and variance $1/\beta$. The relaxation time is $\tau=1/\gamma$. In
superstatistics, one considers a Brownian particle moving through an
environment with a slowly fluctuating temperature
field whose changes take place
on a typical time scale
$T$ such that $\tau<<T$. When this inequality holds, the local velocity
distribution of the system can relax to a Gaussian distribution before
the next change of $\beta(t)$ takes place. As such, after a long time,
the stationary velocity distribution $P(v)$ of the particle is just a
superposition of Gaussian distributions weighted with a function $f(\beta)$
\begin{eqnarray}\label{super_ap}
P(v)&\approx&\int_{\beta_\textrm{min}}^{\beta_\textrm{max}} d\beta
f(\beta)\sqrt{\frac{m\beta}{2\pi}}\exp\left(-\frac12m\beta v^2\right).
\end{eqnarray}
This $f(\beta)$ is the probability density to observe some value of $\beta$.
Depending on the properties of $f(\beta)$, different results for the stationary
velocity distribution $P(v)$
will occur \cite{touchette}, e.g., power-laws or stretched exponentials.
In \cite{unicla,vdsbeck1} a method was introduced to determine $f(\beta)$,
given an experimental time series. The method extracts the main
superstatistical parameters out of a given data set and examines the
validity of the superstatistical model assumptions. Depending on the system
under study, one can obtain different results for $f(\beta)$.
We will first briefly outline this method to extract $f(\beta)$ from the data.

The starting point is a discrete time series $v$ containing $n$ data points.
Using the definitions proposed in \cite{unicla,vdsbeck1}, one can
estimate values for the two different time scales $\tau$ and $T$. In case
that the inequality $\tau/T<<1$ holds, one proceeds by dividing the time series
$v$ in $N$ different time slices of length $t$ with $N=\lfloor n/t\rfloor$,
where $\lfloor x\rfloor$ means rounding the value of $x$ to the nearest lower
integer. Then, one calculates the variance of $v$ in each of these
sub-intervals.
The inverse of the variance is an estimator for the value of $\beta$.
As such, one obtains a new series containing $N$ points which are denoted as
$\beta_i$ with $i=1,2,\ldots N$. Within this assumption, the distribution
$P(v)$ is approximated by
\begin{eqnarray}\label{Hu_1}
P(v)\approx\frac1N\sum_{i=1}^N\sqrt\frac{\beta_{i}}{2\pi}e^{-\frac12\beta_{i}
v^2}.
\end{eqnarray}
When $N$ is large enough, one can replace expression (\ref{Hu_1}) by
(\ref{super_ap}) with $f(\beta)$ being the probability density that the
value of the inverse variance in a randomly chosen time slice of length
$T$ equals $\beta$. Notice that the superstatistical approach includes
two approximations. In the first step, one assumes the existence of two time
scales $\tau$ and $T$ such that in every time slice 'local' equilibrium is
reached. Then, $P(v)$ can be approximated by (\ref{Hu_1}). In the second
step one assumes the existence of a distribution $f(\beta)$ replacing the
summation in expression (\ref{Hu_1}) by an integral. Then, $P(v)$ can be
approximated by (\ref{super_ap}).

\section{Application to wind speed fluctuations}

\begin{figure}
\includegraphics[width=0.48\textwidth]{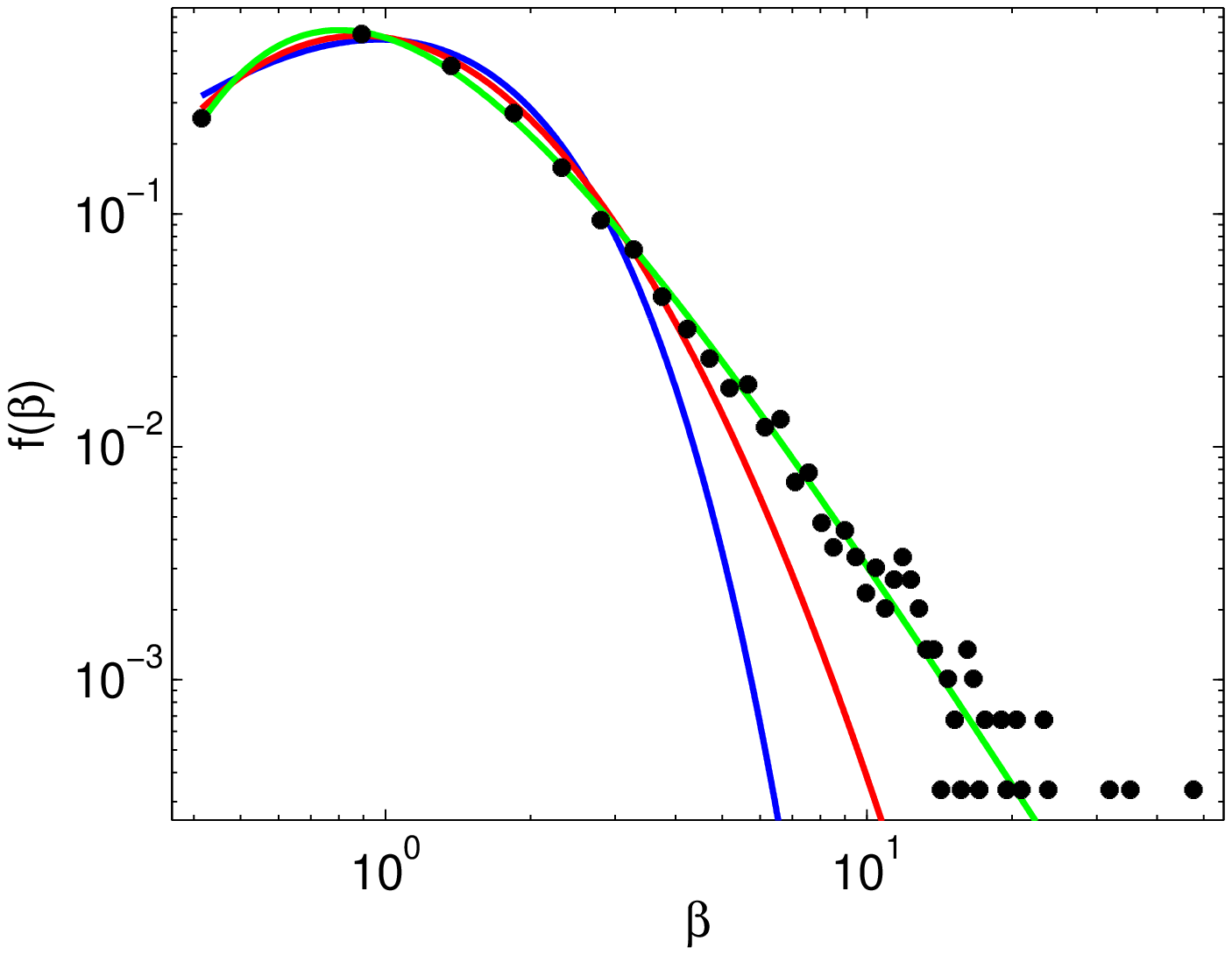}
\includegraphics[width=0.48\textwidth]{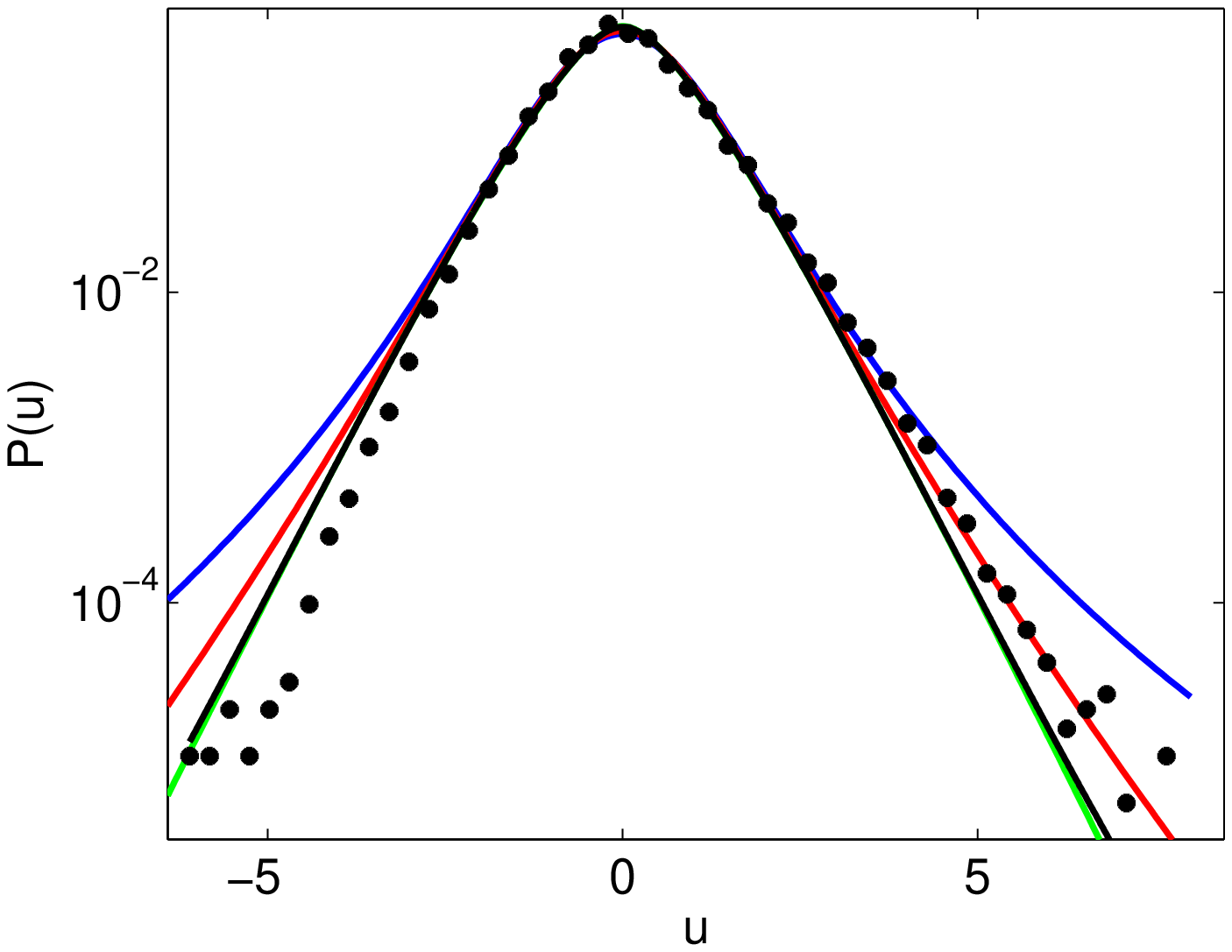}
\caption{\label{fig1}(Colour online) LEFT: Plot of the
empirical distribution $f(\beta)$ (dots)
extracted from the wind data
and the best fit to a lognormal distribution (red), gamma distribution
(blue) and an inverse gamma distribution (green). RIGHT: Plot of the empirical
 distribution $P(u)$ together with the first
superstatistical approximation (\ref{Hu_1}) (black line). Also the second
approximation  (\ref{super_ap}) is shown,
where $f(\beta)$ is given by the lognormal distribution (red), the gamma
distribution (blue) and the inverse gamma distribution (green).}
\end{figure}

In the following,
as a working example,
we apply our method to
an experimentally measured
time series of the horizontal component $v(t)$ of wind speeds recorded
af the Lammefjord site
at a height of 10m (cup 1) during the year 1987 \cite{wind-data}.
The incremental distribution of
$u_\delta(t):=v(t+\delta)-v(t)$ exhibits non-Gaussian behavior that can be
 modelled using superstatistics \cite{laub}. As an illustration, we calculated
the superstatistical approximation of the incremental distribution
($\delta=8$) of the windspeeds at day $191$. The measuring frequency is
$8$Hz. This means that the number of data points is
$n=24\times3600\times8\approx7\times10^5$. The two different time scales
extracted from the data are $\tau\approx 4.1$ and $T \approx 112$,
obtained by using similar techniques as in \cite{unicla,vdsbeck1}.
This illustrates that the data set shows clear time scale separation, in
agreement with the results of \cite{kantz}.
Knowing the value of $T$, one can construct the distribution $f(\beta)$ and try
to approximate this histogram with some well-known distributions
such as the gamma distribution, the lognormal distribution or the inverse
gamma distribution. The latter one is given by
\begin{eqnarray}\label{fbeta_infg}
f(\beta)&=&\frac{\theta^\alpha}{\Gamma(\alpha)}\beta^{-\alpha-1}e^{-\theta/\beta
}.
\end{eqnarray}
Here $\alpha$ and $\theta$ are parameters.
The relevance of the
above three distributions was motivated in \cite{unicla}.
The results of our calculations are shown in figure \ref{fig1}.
There is an excellent agreement between the histogram extracted from the
data and the inverse gamma distribution. The right part of the
figure also shows the
empirical distribution $P(u)$ together with the first and second
approximation of superstatistics
given by eqs. (\ref{Hu_1}) and (\ref{super_ap}),
respectively. The first approximation of superstatistics well
models the fat tails of the empirical distribution $P(u)$, see Figure
\ref{fig1}.
The excellent agreement between $P(u)$ and expression (\ref{super_ap}), with
$f(\beta)$ being the inverse gamma distribution, gives further evidence that
this latter distribution is well able to represent the fluctuations of $\beta$.
Summarizing, our simple superstatististical model discussed so
far is a good first-order
approximation of the process of windspeeds increments.

\begin{figure}
\includegraphics[width=0.48\textwidth]{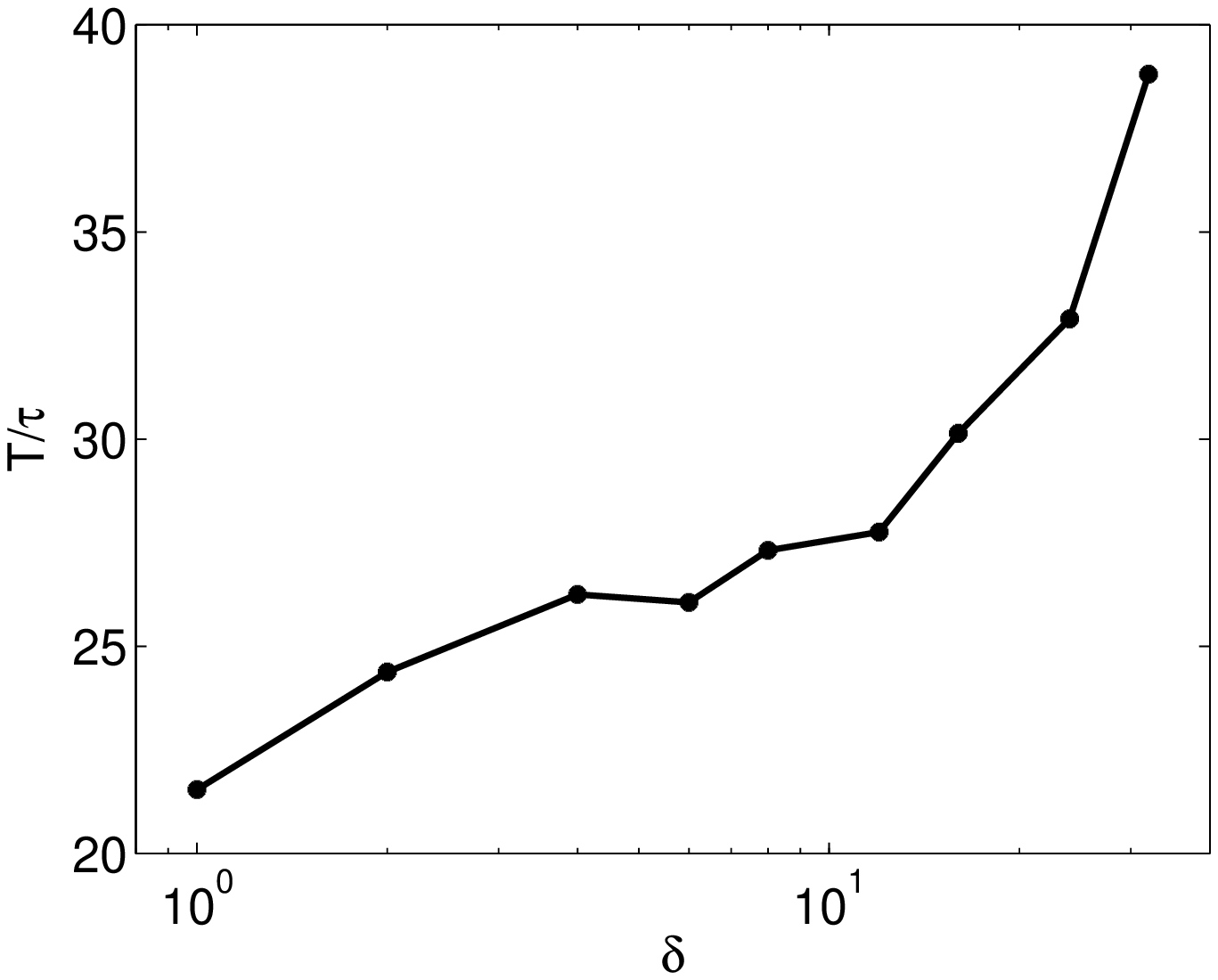}
\includegraphics[width=0.48\textwidth]{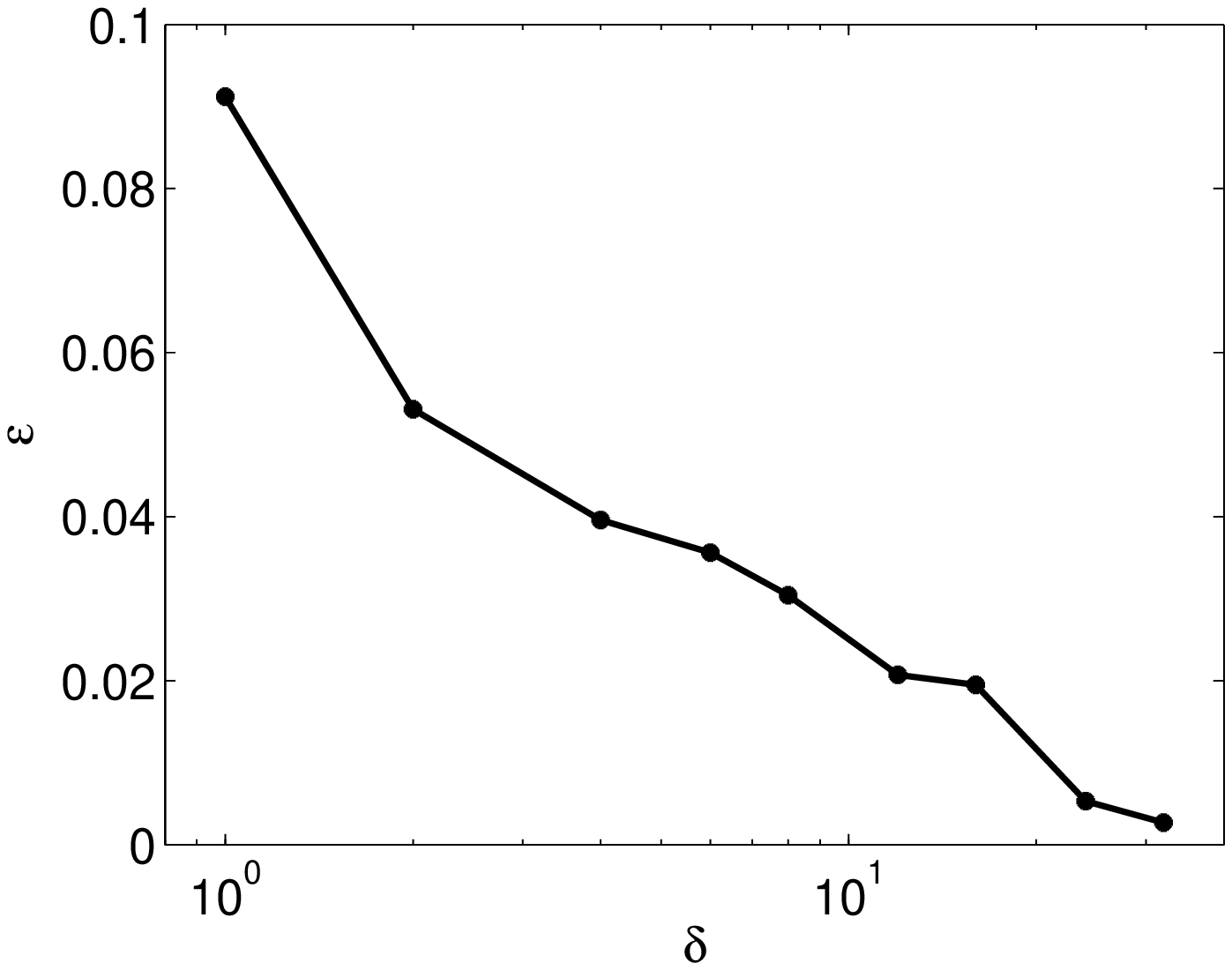}
\caption{\label{fig2}(Colour online) LEFT: Ratio of the
time scales $T$ and $\tau$ as extracted from the data. RIGHT: The parameter
$\epsilon$, defined in \cite{vdsbeck1}, measures how good a superstatistical
model fits a given data set. Ideal superstatistics is corresponding to
$\epsilon =0$.}
\end{figure}

Figure \ref{fig2} shows the extracted time scale ratio $T/\tau$ as a function of
$\delta$.
Apparently,
for small $\delta$ the time scale separation between $T$ and $\tau$
is less pronounced. The right part of the figure also shows a parameter
$\epsilon$
as a function of $\delta$. This $\epsilon$ was defined in \cite{vdsbeck1} as a
measure
of quality of the superstatistical approximation. The smaller $\epsilon$, the
better
the superstatistical model assumptions are satisfied for the given time series.

\section{Superstatistics and overdamped motion}
We will now generalize the Langevin approach to superstatistics.
We start from the
following set of equations
\begin{eqnarray}\label{xv_lan}
\frac d{dt}v(t)&=&-\gamma v(t)+\frac1mF(x)+\sqrt{\frac{2\gamma}m
\frac1{\beta(x)}}g(t)
\cr
\frac d{dt}x(t)&=&v(t),
\end{eqnarray}
with $F(x)=-\partial V(x)/\partial x$. Here $V(x)$ is a confining potential
like, e.g., a harmonic potential, and $x$ is a position
variable. Crucial for our approach
is that $\beta$ depends on $x$ but not on the velocity $v$.
We are interested in the stationary velocity distribution $P(v)$. It is well
known that in case $\beta(x)=\beta$ is a constant, this distribution becomes
a Gaussian
distribution
\begin{eqnarray}\label{ggg}
P(v)=\sqrt{\frac{m\beta}{2\pi}}\exp\left(-\frac12m\beta v^2\right).
\end{eqnarray}
However, we are interested in the general problem where the temperature
is position dependent. Then, in order to obtain an exact expression for $P(v)$,
one has to solve the corresponding stationary Fokker-Planck equation for the
joint distribution $P(v,x)$,
\begin{eqnarray}\label{full_fp}
0&=&\left[\frac{\partial }{\partial v}\left(\gamma
v-\frac1mF(x)\right)-v\frac{\partial}{\partial x}+\frac\gamma m
\frac1{\beta(x)}\frac{\partial^2 }{\partial^2v}\right]P(v,x)
\end{eqnarray}
and integrate out the dependence of the variable $x$
\begin{eqnarray}
P(v)&=&\int_{x_\textrm{min}}^{x_\textrm{max}} dx P(v,x).
\end{eqnarray}
The problem with this procedure is that usually it is not possible to obtain
an analytical expression for $P(v,x)$. However, in order to make the
connection with the superstatistical approach outlined in the
previous section, it is not necessary to have the exact form of $P(v,x)$.
It is sufficient to study (\ref{xv_lan}) in the high friction limit, i.e.,
$\gamma\rightarrow\infty$. Taking this overdamped limit of (\ref{xv_lan})
results in a stochastic differential equation in the variable $x(t)$ only:
\begin{eqnarray}
\frac
d{dt}x(t)&=&\frac1{m\gamma}F(x)+\sqrt{\frac2{m\gamma}\frac1{\beta(x)}}g(t)
\end{eqnarray}
The associated Fokker-Planck equation for the stationary distribution
$P_o(x)$ becomes
\begin{eqnarray}\label{over_fp}
0&=&-\frac{\partial }{\partial x}F(x)P_o(x)+
\frac{\partial^2 }{\partial^2x}\frac1{\beta(x)}P_o(x)
\end{eqnarray}
(we used Ito's interpretation).
The lower index $\ _o$ is used to denote the
overdamped limit. Equation (\ref{over_fp}) can be solved analytically
\begin{eqnarray}
P_o(x)&=&Z^{-1}\beta(x)\exp\left(\int dx F(x)\beta(x)\right),
\end{eqnarray}
where $Z$ is a normalization constant. Taking the overdamped limit physically
means that the velocity will thermalise very quickly. As a consequence, for
large values of $\gamma$, the solution $P(v,x)$ of the complete Fokker-Planck
equation (\ref{full_fp}) is approximated by \cite{stolo}
\begin{eqnarray}\label{dist_over}
P(v,x)\approx P_o(v,x)=P_o(x)P_o(v|x).
\end{eqnarray}
Here $P_o(v|x)$ denotes the conditional distribution
of $v$ given the local value of the slow variable $x$. This is just a
Gaussian distribution (\ref{ggg}) with $\beta$ replaced by $\beta(x)$. As a
consequence, one obtains the following approximation for the distribution
of interest:
\begin{eqnarray}\label{faaa}
P(v)\approx \int_{x_\textrm{min}}^{x_\textrm{max}} dx
P_0(v,x)=\int_{x_\textrm{min}}^{x_\textrm{max}}
dxP_o(x)\sqrt{\frac{m\beta(x)}{2\pi}}\exp\left(-\frac12m\beta(x)v^2\right).
\end{eqnarray}
After a change of variables, one ends up with (\ref{super_ap}) in which the
superstatistical distribution $f(\beta)$ is associated with
\begin{eqnarray}
\left(\frac{d\beta(x)}{dx}\right)^{-1}P_o(\beta(x)).
\end{eqnarray}
This shows that the superstatistical approximation is formally equivalent to
taking the high friction limit of a stochastic differential equations of
type (\ref{xv_lan}). The latter model describes the dynamics of a
Brownian particle in an inhomogeneous heat bath in the presence of a
confining potential.

\section{Example}\label{exwin}
To illustrate the theoretical considerations of the previous section, we study
an explicit example here. Assume for simplicity that the temperature is linear
in the position $x$. In order to ensure that the temperature remains positive we
introduce an arbitrary small positive constant $a$ and define
\begin{eqnarray}\label{defFV}
\beta(x)=\frac1{|x|+a},&&V(x)=\frac\theta2\left[|x|+a-\frac\alpha\theta\right]
^2.
\end{eqnarray}
With this choice one obtains for the velocity distribution in the overdamped
limit (\ref{faaa})
\begin{eqnarray}
P(v)&\approx&Z^{-1}\int_{-\infty}^{+\infty}
dx\beta(x)\exp\left(-\int dx' \frac{\partial V(x')}{\partial
x'}\beta(x')\right)\sqrt{\frac{
m\beta(x)}{2\pi}}\exp\left(-\frac12m\beta(x) v^2\right)
\cr
&=&Z^{-1}\int_{-\infty}^{+\infty}
dx\exp\left((-\alpha+1)\ln\beta(x)-\frac\theta{\beta(x)}\right)\sqrt{\frac{
m\beta(x)}{2\pi}}\exp\left(-\frac12m\beta(x) v^2\right)
\cr
&=&Z^{-1}\int_{0}^{1/a} d\beta\
\beta^{-\alpha-1}\exp\left(-\frac\theta\beta\right)\sqrt{\frac{m\beta}{2\pi}}
\exp\left(-\frac12m\beta v^2\right)
\end{eqnarray}
where we absorbed constants in the normalization $Z$. In the limit
$a\rightarrow0$, this integral simplifies to
\begin{eqnarray}\label{pv_invgam}
P(v)&\approx&\frac1{\Gamma(\alpha)}\sqrt{\frac{2\theta}\pi}\left(\sqrt{
\frac\theta2}|v|\right)^{\alpha-1/2}\textrm{K}_{\alpha-1/2}\left(\sqrt{2\theta}
|v|\right)
\end{eqnarray}
where $\textrm{K}_{\nu}(x)$ is the modified Bessel function of the second kind.
Notice that $f(\beta)$
becomes an inverse gamma distribution in this limit. This particular example
connects the results of sections III and IV because it shows that the
theoretical model of section IV together with the choice (\ref{defFV}) for
the temperature and the potential reproduces the observed empirical
distributions
discussed in section III. However, this is only one possibility. In principle,
for every choice of $\beta(x)$, a possible candidate for the force that results
in an inverse gamma distribution for $f(\beta)$ can be calculated by evaluating
the following expression
\begin{eqnarray}\label{dfadaa}
F(x)\beta(x)&=&-(\alpha+2)\frac{\beta'(x)}{\beta(x)}-\theta\frac{\beta'(x)}{
\beta^2(x)}+\frac{\beta''(x)}{\beta'(x)}.
\end{eqnarray}
Where the accent means differentiating with respect to $x$. The reason to chose
the particular form of $T(x)$ and $V(x)$ as in (\ref{defFV}) is that this is
probably the easiest choice that both fulfills equation (\ref{dfadaa}) and is
physically meaningful. The latter is also the reason that we
introduced the constant $a$ in the expression for the temperature (\ref{defFV}).
This is to ensure that
the particle can pass the origin in the overdamped limit. This problem
disappears if one takes finite friction into account. Therefore, it is
appropriate to
restrict our analysis to the special case $a\rightarrow0$.

\section{Skewness}
\begin{figure}
\includegraphics[width=0.48\textwidth]{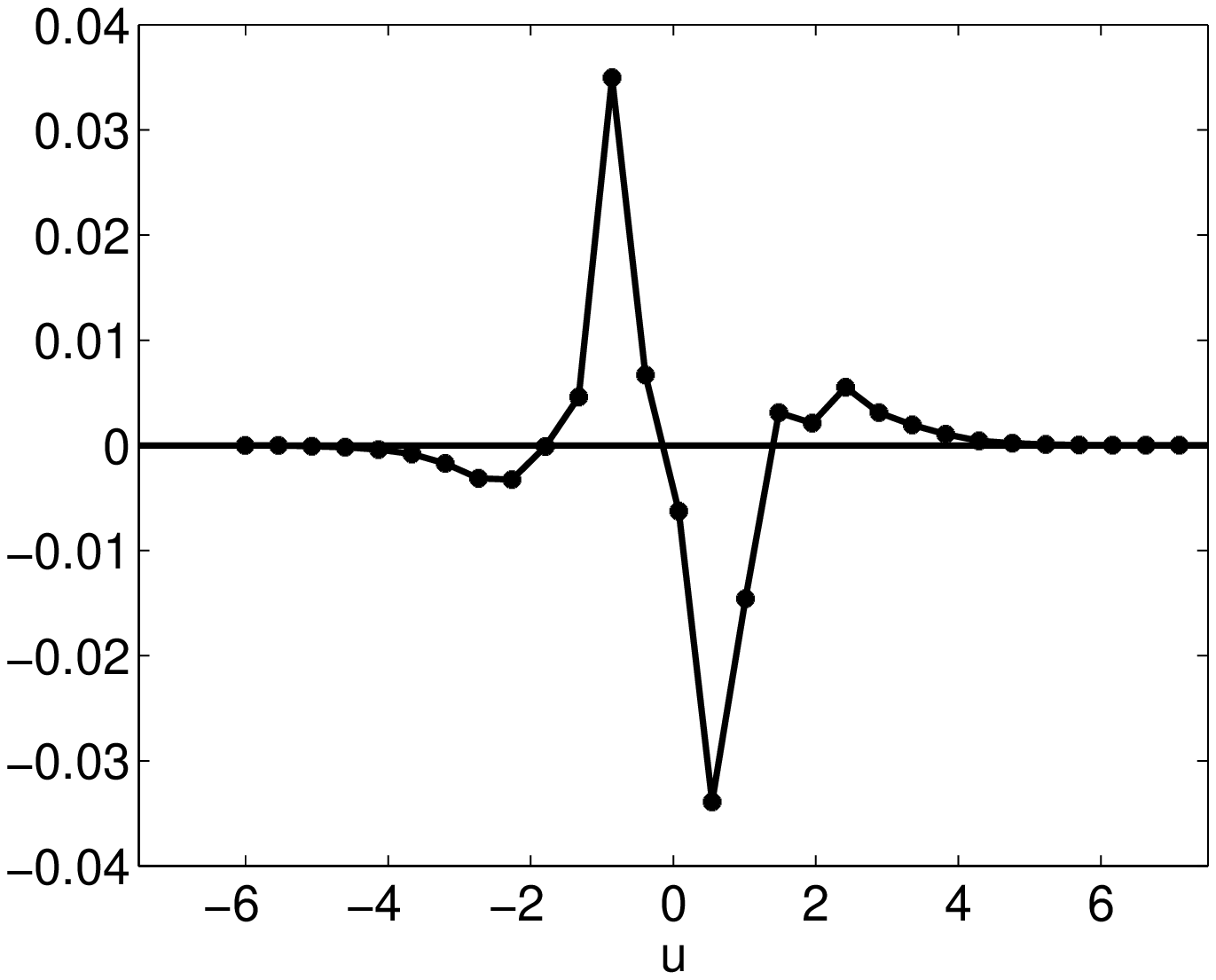}
\includegraphics[width=0.48\textwidth]{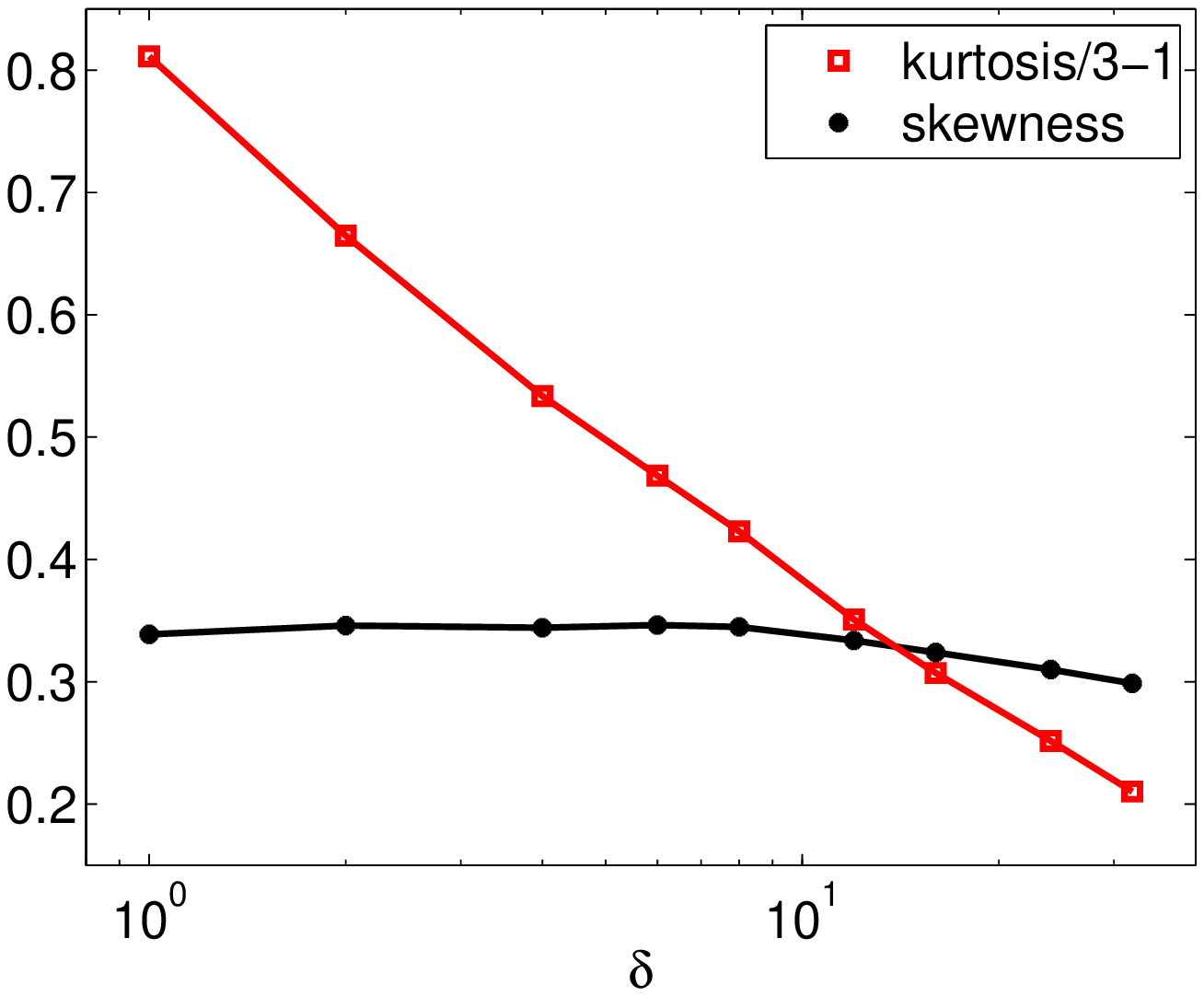}
\caption{\label{fig3}(Colour online) LEFT: The difference between the empirical
distribution $P(u)$ and the second approximation of
 superstatistics. RIGHT: Kurtosis and skewness of the empirical
distribution $P(u)$ as a function of $\delta$.}
\end{figure}
We showed in section III that the superstatistical approach is well suited to
describe the fat tails that are exhibited in the wind speed distribution under
study.
However, it is well known, and clearly visible in figure \ref{fig1}, that the
distribution $P(u)$ is slightly skewed (with vanishing first moment). Clearly,
the
superstatistical distribution (\ref{super_ap}) is symmetric and can not
represent the skewness that occurs in the experimental data. Figure \ref{fig3}
 (right part) shows the skewness and kurtosis as a function of
$\delta$ for the data set of measured velocity increments $u_\delta(t)$. It is
possible to model the kurtosis with the superstatistical approach described up
till now, however it is impossible to model the skewness.

As mentioned before, the complete Fokker-Planck equation (\ref{full_fp})
is usually not analytically solvable. However, relations between the
moments of the distribution $P(v,x)$ can be obtained, using the Fokker-Planck
operator $L$. One has
\begin{eqnarray}\label{fp_operator}
\langle Lv^ix^j\rangle=0&\textrm{for}&i,j=0,1,2,\ldots \textrm{with} \,
 L:=\left(\gamma v-\frac1mF(x)\right)\frac{\partial }{\partial
v}-v\frac{\partial}{\partial x}-\frac\gamma m \frac1{\beta(x)}\frac{\partial^2
}{\partial^2v}.
\end{eqnarray}
Applying this with $i=0,j=1$ results in $\langle v\rangle=0$, while depending on
the particular choices of $F(x)$ and $\beta(x)$ one can
obtain $\langle v^3\rangle\neq0$. This means that the stationary velocity
distribution $P(v)$ can have non-vanishing skewness, while
at the same time it has a vanishing first moment. However notice that the
overdamped approximation (\ref{dist_over}) of $P(v,x)$ is symmetric in $v$,
regardless of the choice of $\beta(x)$. This means that in order to introduce
skewness, one has to go beyond the overdamped limit. Perturbation expansions of
$P(v,x)$ around the overdamped limit can be performed \cite{stolo}. The first
order correction is
\begin{eqnarray}
P(v,x)&=&\sqrt{\frac{m\beta(x)}{2\pi}}
\exp\left(-\frac12m\beta(x)v^2\right)P_o(x)\left(1+\frac16\frac1\gamma
v\left[m\beta(x)v^2-3\right]\frac1{\beta(x)}\frac{d\beta(x)}{dx}\right).
\end{eqnarray}
Notice that the contribution of the first order correction vanishes for $\langle
v^i\rangle$ with $i=0,1,2$. The third order moment becomes
\begin{eqnarray}\label{ddd}
\langle v^3\rangle=\int_{-\infty}^{+\infty}dv\int_{-\infty}^{+\infty}dx
v^3P(v,x)&=&\frac1\gamma\frac1{m^2}\int_{-\infty}^{+\infty}dxP_o(x)
\frac1{\beta^3(x)}\frac{d\beta(x)}{dx}.
\end{eqnarray}
This shows that the theoretical model introduced in section IV can be used to
systematically describe the experimental data. The superstatistical
approximation (the overdamped limit) is used to model the fat tails (the
kurtosis) of the empirical distribution. By going beyond the overdamped limit,
skewness is naturally introduced in the theoretical distribution.

\section{Discussion}
In previous papers dealing with superstatistics, the superstatistical concept
was completely probabilistic in
nature. One was facing the question what
the relevant temperature distributions $f(\beta)$ are. In this paper, we 
have shown that the superstatistical model is equivalent to a dynamical process
with a position
dependent temperature field in the overdamped limit. This 
opens up the possibility to relate the choice of $f(\beta)$ to a
concrete question with a clear physical interpretation, namely
what are appropriate choices for the potential and the 
temperature fields.

One such example was
studied in this paper. If one assumes a temperature field that is linear in the
position, in combination with basically a harmonic potential, one ends up with
an
inverse gamma distribution for $f(\beta)$. This distribution is indeed
observed in empirical data of wind speed fluctuations. We emphasize that other
choices for temperature field and potential are of course possible. It is an
interesting topic for further research to examine other physically meaningfull
combinations of $\beta(x)$ and $V(x)$. One can also try to extract the position
dependence of $\beta(x)$ and $V(x)$ immediately out of empirical data, see e.g.
\cite{chin_eco} where such analysis is performed for financial time series.

Finally, we showed that the theoretical approach to superstatistics presented in
this paper naturally leads to skewed probability distributions if one goes
beyond the overdamped limit. Slightly skewed distributions are observed in
several scientific fields for various variables, e.g., wind speed fluctuations
(see
section VI), log returns of prices in the stock market \cite{eco_sk}, and
velocity
increments of turbulent flows \cite{turb_sk}. Apart from the observed skewness
in
the data, these three examples have in common that the symmetric part of the
empirical distribution can be well approximated by (\ref{super_ap}) with
different expressions for $f(\beta)$, see, e.g., the present paper and
\cite{vdsbeck1}. In the theoretical model developed in this paper, expression
(\ref{super_ap}) is the zeroth-order approximation of the true superstatistical
distribution which is obtained by fully taking into account the position
dependence of
$\beta(x)$ and $V(x)$. Therefore, the current approach is very
promising, since one can now go beyond (\ref{super_ap}) and try to calculate
successive higher-order corrections in order to describe all the essential
features of an empirical distribution at hand. Apart from the aforementioned
examples, also dense granular flows are interesting experimental systems from a
superstatistical point of view, because position dependent temperature profiles
can be
estimated for these kinds of systems. Also, it has been shown that the particle
velocity
distribution shows skewness \cite{gran_flow}.

\end{document}